\documentclass[superscriptaddress,twocolumn, preprintnumbers,nofootinbib,amsmath,amssymb]{revtex4-1}
\pdfoutput=1
\usepackage{graphicx}% Include figure files
\usepackage{dcolumn}% Align table columns on decimal point
\usepackage{bm}% bold math
\usepackage{soul}
\usepackage{xcolor}
\usepackage[hidelinks]{hyperref} %[linktocpage=true]            

\newcommand\ee{\end{equation}}
\newcommand\be{\begin{equation}}
\newcommand\eea{\end{eqnarray}}
\newcommand\bea{\begin{eqnarray}}

\newcommand\di{\partial}

\newcommand\comment[1]{}
\usepackage[hidelinks]{hyperref} %[linktocpage=true]
\comment{
\hypersetup{
colorlinks=true,
citecolor=DarkBlue,
linkcolor=DarkBlue,
urlcolor=DarkBlue,
}}

\begin{document}

\title{The Quasinormal Modes of Quasinormal Modes}

\author{Mehrdad Mirbabayi}
\affiliation{International Centre for Theoretical Physics, Trieste, Italy}

\begin{abstract}

Modifying black hole horizon can drastically change the spectrum of quasinormal modes. But if the modification is close enough to the horizon the early ringdown signal remains almost unaltered, and well described by the quasinormal modes of the original GR solution. I show how the original quasinormal modes emerge in the sum over the new modes. 

\end{abstract}
\maketitle
\noindent
Einstein theory of gravity has a unique prediction for how perturbations around black holes relax. This relaxation process that is called the ringdown phase is accompanied by the emission of gravitational waves. Modifications of the Einstein theory or environmental effects generically change those predictions, giving us the opportunity to test them against observations, see e.g. \cite{Ligo}. 

Traditionally, the ringdown signal is characterized by the spectrum of the quasinormal modes. These are the eigenmodes of the wave equation for linearized perturbations with out-going boundary condition at infinity and in-going condition at the horizon (when there is one). They appear as the complex poles of the retarded Green's function in the frequency space. For any stable system the quasinormal modes decay with time and hence they give a useful description of the relaxation process. See \cite{Kokkotas,Konoplya} for reviews. 

Modifications of the Einstein theory or the background black hole solution can drastically change the quasinormal mode spectrum, nevertheless there are examples in which the ringdown signal remains almost the same for a long time. The goal of this note is to explain how this happens via a simple model.

It is a straight consequence of causality that at early times black hole ringdown signal has to be identical to what GR predicts, provided any modification to the black hole solution is close enough to the horizon. In particular, if the linearized perturbations experience the same effective potential as in the Schwarzschild geometry until they reach large negative values of tortoise coordinate
\be
r+r_g\log(r-r_g)\sim -a \ll -r_g,
\ee
the effect of the modification is observed as ``echoes'', arriving with a delay of order $a$ with respect to the early ringdown signal \cite{Cardoso}. See \cite{Cardoso2,Price,Nakano,Volkel,Mark,Bueno} for further discussions of this scenario.\footnote{Echoes can also be caused by the reflection of the waves from matter distribution far from the horizon. See e.g. \cite{Barausse,Konoplya2}. The focus here will be on modifications near the horizon, though a similar argument is expected to apply to the latter case.}

To model this situation consider a $1+1-d$ wave equation 
\be
[\di_t^2 - \di_x^2 + V_1(x)+V_2(x)]\psi(t,x)=0
\ee
where $V_1(x)$ and $V_2(x)$ are {\em hard barriers} localized, respectively, at $a_1$ and $a_2<a_1$. What I mean by {\em hard barrier} is that in isolation $V_1$ and $V_2$ have a spectrum of quasinormal modes with characteristic frequency $|\omega| \gg 1/(a_1-a_2)$. This implies that for an initial perturbation at $x_0>a_1$, the observer at $x > x_0$ receives an {\em early} response after 
\be
t_1\equiv x+x_0-2a_1,
\ee
which is characteristic of $V_1(x)$. For instance, if $V_1$ is the effective potential for perturbations on Schwarzschild background with gravitational radius $r_g$, this response has an exponentially falling profile $\exp(-\lambda (t-t_1))$, with $\lambda \sim 1/r_g$. The existence of the second barrier would matter only after 
\be
t_2\equiv x+x_0-2a_2,
\ee
and the requirement of hardness is $(t_2-t_1)\gg r_g$.

In the very same regime, the spectrum of the quasinormal modes of the double-peak potential is completely different. Indeed, they have a simple interpretation as a set of almost stable bound-states, which are trapped between the two barriers. They have approximately equidistant real parts
\be\label{Re}
{\rm Re~} \omega_n \simeq \frac{n\pi}{a_1 -a_2},
\ee
with small imaginary part ${\rm Im ~} \omega_n = \mathcal{O}(1/({a_1-a_2}))$, resulting from the leakage through the two hard barriers. The question is how the early signal matches the non-existing quasinormal modes of an isolated $V_1$ potential.

For a general potential $V(x)$, the response at large positive $x$ to an in-falling initial perturbation $\psi_0=\delta(t+x-x_0)$, is
\be\label{psi}
\psi(t,x)= \int_{-\infty}^\infty \frac{d\omega}{2\pi} R(\omega) e^{-i\omega(t-(x+x_0))},
\ee
where $R(\omega)$ is the reflection coefficient. $R(\omega)$ and $T(\omega)$, the transmission coefficient, are defined in terms of $x\to \infty$ behavior of the solution of 
\be 
[-\di_x^2+V(x)]\phi(x) =\omega^2 \phi(x),
\ee
with the boundary condition $\phi(x\to -\infty) = e^{-i\omega x}$:
\be
\phi(x\to \infty) = \frac{1}{T(\omega)} e^{-i\omega x} + \frac{R(\omega)}{T(\omega)} e^{i\omega x}.
\ee
For the double-peak potential, they can be easily related to $R_{1,2}(\omega)$ and $T_{1,2}(\omega)$, the reflection and transmission coefficients from the first and the second barriers. We get (dropping the $\omega$ arguments for brevity)
\be\label{R/T}
\frac{R}{T}=\left(\frac{R_1}{T_1 T_2}e^{-2i\omega a_1}+\frac{R_2}{\bar T_1T_2} e^{-2i\omega a_2}\right)
\ee
and
\be\label{T}
\frac{1}{T} =\left(\frac{1}{T_1 T_2}+\frac{\bar R_1 R_2}{\bar T_1T_2} e^{2i\omega (a_1-a_2)}\right),
\ee
where $\bar R_1(\omega)=R_1(-\omega)$ and $\bar T_1(\omega)=T_1(-\omega)$.
\comment{
For instance, in our explicit example of two delta-function peaks with equal strength $v$,
\be
R_1=R_2=\frac{-iv/2\omega}{1+iv/2\omega},\qquad T_1=T_2 = \frac{1}{1+iv/2\omega}.
\ee
}

Substituting $R(\omega)$ from \eqref{R/T} and \eqref{T} in \eqref{psi}, we can evaluate the $\omega$ integral using the residue theorem. There is no singularity in the upper-half plane. At very early times, the integration contour is closed in the upper-half plane and we get $0$. For $t_1<t<t_2$, the first term in \eqref{R/T} becomes relevant,
\be\label{psi1}
\psi(t,x)= \int^\infty_{-\infty} d\omega \frac{R_1 e^{-i\omega(t-(x+x_0-2a_1))}}
{2\pi [1+T_1 R_2 (\bar R_1/\bar T_1) \exp(2i\omega(a_1-a_2))]}
.
\ee
Closing the contour in the lower-half plane, we pick up the residues of the poles, which are solutions $\{\omega_n\}$ to 
\be\label{F}
F(\omega)=1+T_1 R_2 (\bar R_1/\bar T_1) e^{2i\omega(a_1-a_2)}=0,
\ee
while since $|R_1|^2+|T_1|^2=1$, the poles of $R_1$ in the numerator of \eqref{psi1} cancel with the poles of $T_1$ in the denominator. Hence, one obtains
\be\label{sum}
\psi(t,x) \simeq \sum_n \frac{R_1(\omega_n)}{2(a_1 - a_2)}e^{-i\omega_n(t-(x+x_0-2a_1))},
\ee
where I used the hard-barrier assumption to approximate
\be\label{dF}
\left.\frac{dF}{d\omega}\right|_{\omega_n}\simeq-2i (a_1-a_2).
\ee
(Since $a_1-a_2$ is the largest length scale in the problem, the dominant term in $dF/d\omega$ comes from taking the derivative of $e^{2i\omega(a_1-a_2)}$. The result can then be simplified using $F(\omega_n)=0$.)

Note from \eqref{F} that $|{\rm Im ~} \omega_n| \sim 1/(a_1-a_2)$, which is much smaller than the characteristic time-scales of interest. Neglecting ${\rm Im~ } \omega_n$ and using the regular spacing \eqref{Re} of the ${\rm Re~} \omega_n$, the sum \eqref{sum} can be approximated by the following integral
\be
\psi(t,x)\simeq \int_{-\infty}^\infty \frac{d\omega}{2\pi} R_1(\omega) e^{-i\omega(t-(x+x_0-2a_1))}.
\ee
This is the response of an isolated $V_1$ potential, localized near $a_1$. 

%In our delta-function toy example, it reduces by the residue theorem to a single quasinormal mode at $\omega = -iv/2$.

Therefore, as long as there is a hierarchy between the characteristic frequency of two separate features in the potential and their separation $a_1 - a_2$, the above argument shows that at early times the response is well-described by the quasinormal modes of the closest feature, as it should.

\vspace{0.3cm}
\noindent
{\em Acknowledgments.---} I thank Vitor Cardoso and Paolo Creminelli for stimulating discussions. This work was partially supported by the Simons Foundation Origins of the Universe program (Modern Inflationary Cosmology collaboration).

%\bibliographystyle{aipauth4-1}
%\bibliography{refs}

\begin{thebibliography}{99}
\bibitem{Ligo} 
  B.~P.~Abbott {\it et al.} [LIGO Scientific and Virgo Collaborations],
  ``Tests of general relativity with GW150914,''
  Phys.\ Rev.\ Lett.\  {\bf 116}, no. 22, 221101 (2016)
  Erratum: [Phys.\ Rev.\ Lett.\  {\bf 121}, no. 12, 129902 (2018)]
  doi:10.1103/PhysRevLett.116.221101, 10.1103/PhysRevLett.121.129902
  [arXiv:1602.03841 [gr-qc]].
\bibitem{Kokkotas} 
  K.~D.~Kokkotas and B.~G.~Schmidt,
  ``Quasinormal modes of stars and black holes,''
  Living Rev.\ Rel.\  {\bf 2}, 2 (1999)
  doi:10.12942/lrr-1999-2
  [gr-qc/9909058].
\bibitem{Konoplya} 
  R.~A.~Konoplya and A.~Zhidenko,
  ``Quasinormal modes of black holes: From astrophysics to string theory,''
  Rev.\ Mod.\ Phys.\  {\bf 83}, 793 (2011)
  doi:10.1103/RevModPhys.83.793
  [arXiv:1102.4014 [gr-qc]].
\bibitem{Cardoso} 
  V.~Cardoso, E.~Franzin and P.~Pani,
  ``Is the gravitational-wave ringdown a probe of the event horizon?,''
  Phys.\ Rev.\ Lett.\  {\bf 116}, no. 17, 171101 (2016)
  Erratum: [Phys.\ Rev.\ Lett.\  {\bf 117}, no. 8, 089902 (2016)]
  doi:10.1103/PhysRevLett.117.089902, 10.1103/PhysRevLett.116.171101
  [arXiv:1602.07309 [gr-qc]].
%\cite{Cardoso:2016oxy}
\bibitem{Cardoso2} 
  V.~Cardoso, S.~Hopper, C.~F.~B.~Macedo, C.~Palenzuela and P.~Pani,
  ``Gravitational-wave signatures of exotic compact objects and of quantum corrections at the horizon scale,''
  Phys.\ Rev.\ D {\bf 94}, no. 8, 084031 (2016)
  doi:10.1103/PhysRevD.94.084031
  [arXiv:1608.08637 [gr-qc]].
  %%CITATION = doi:10.1103/PhysRevD.94.084031;%%
  %113 citations counted in INSPIRE as of 07 Jan 2019

%\cite{Price:2017cjr}
\bibitem{Price}
  R.~H.~Price and G.~Khanna,
  ``Gravitational wave sources: reflections and echoes,''
  Class.\ Quant.\ Grav.\  {\bf 34}, no. 22, 225005 (2017)
  doi:10.1088/1361-6382/aa8f29
  [arXiv:1702.04833 [gr-qc]].
  %%CITATION = doi:10.1088/1361-6382/aa8f29;%%
  %28 citations counted in INSPIRE as of 07 Jan 2019

%\cite{Nakano:2017fvh}
\bibitem{Nakano}  
  H.~Nakano, N.~Sago, H.~Tagoshi and T.~Tanaka,
  ``Black hole ringdown echoes and howls,''
  PTEP {\bf 2017}, no. 7, 071E01 (2017)
  doi:10.1093/ptep/ptx093
  [arXiv:1704.07175 [gr-qc]].
  %%CITATION = doi:10.1093/ptep/ptx093;%%
  %23 citations counted in INSPIRE as of 07 Jan 2019

%\cite{Volkel:2017kfj}
\bibitem{Volkel} 
  S.~H.~V\"olkel and K.~D.~Kokkotas,
  ``Ultra Compact Stars: Reconstructing the Perturbation Potential,''
  Class.\ Quant.\ Grav.\  {\bf 34}, no. 17, 175015 (2017)
  doi:10.1088/1361-6382/aa82de
  [arXiv:1704.07517 [gr-qc]].
  %%CITATION = doi:10.1088/1361-6382/aa82de;%%
  %15 citations counted in INSPIRE as of 07 Jan 2019
%\cite{Mark:2017dnq}
\bibitem{Mark} 
  Z.~Mark, A.~Zimmerman, S.~M.~Du and Y.~Chen,
  ``A recipe for echoes from exotic compact objects,''
  Phys.\ Rev.\ D {\bf 96}, no. 8, 084002 (2017)
  doi:10.1103/PhysRevD.96.084002
  [arXiv:1706.06155 [gr-qc]].
  %%CITATION = doi:10.1103/PhysRevD.96.084002;%%
  %39 citations counted in INSPIRE as of 07 Jan 2019
%\cite{Bueno:2017hyj}
\bibitem{Bueno} 
  P.~Bueno, P.~A.~Cano, F.~Goelen, T.~Hertog and B.~Vercnocke,
  ``Echoes of Kerr-like wormholes,''
  Phys.\ Rev.\ D {\bf 97}, no. 2, 024040 (2018)
  doi:10.1103/PhysRevD.97.024040
  [arXiv:1711.00391 [gr-qc]].
  %%CITATION = doi:10.1103/PhysRevD.97.024040;%%
  %28 citations counted in INSPIRE as of 07 Jan 2019
\bibitem{Barausse} 
  E.~Barausse, V.~Cardoso and P.~Pani,
  ``Can environmental effects spoil precision gravitational-wave astrophysics?,''
  Phys.\ Rev.\ D {\bf 89}, no. 10, 104059 (2014)
  doi:10.1103/PhysRevD.89.104059
  [arXiv:1404.7149 [gr-qc]].
\bibitem{Konoplya2} 
  R.~A.~Konoplya, Z.~Stuchlík and A.~Zhidenko,
  %``Echoes of compact objects: new physics near the surface and matter at a distance,''
  Phys.\ Rev.\ D {\bf 99}, no. 2, 024007 (2019)
  doi:10.1103/PhysRevD.99.024007
  [arXiv:1810.01295 [gr-qc]].


  
\end{thebibliography}

\end{document}